\documentclass[12pt]{iopart}
\usepackage{amssymb,epsfig,subfigure,rotating}
\usepackage{epsfig,subfigure}
\begin{document}
\title[Synchronisation in two dimensional systems]
{Synchronisation schemes for two dimensional discrete systems}
\author{G. Ambika and K. Ambika}
\address{Department of Physics,
Maharaja's College, Cochin-682011, India.}
\ead{ambika@iucaa.ernet.in,ambikak2002@yahoo.com}

\begin{abstract}
In this work we consider two models of two dimensional discrete
systems subjected to three different types of coupling and analyse 
systematically the performance of each in realising synchronised states.
We find that linear coupling effectively introduce control of chaos along with 
synchronisation, while synchronised  chaotic states are possible with  an 
additive parametric coupling scheme both being equally relevent for specific
applications. The basin leading to synchronisation 
in the initial value plane and the choice of parameter values 
for synchronisation in the parameter plane are isolated in each case.
\end{abstract}

\vspace{2pc}
\maketitle

\section{Introduction}\label{ch.sec1}
The presence of chaos has been
extensively demonstrated in many naturally occurring nonlinear
systems. The capability of the chaotic state to  amplify small
perturbations improves their utility for reaching specific desired
states with very high flexibility and low energy cost. However
there are many situations like power systems where control of
chaos is desirable. This helps in improving a desired behaviour by
making only small time dependant perturbations in a system parameter.
In this context, a key observation is that a chaotic attractor has
embedded within it an infinite set of unstable periodic orbits and
the control mechanism aims at stabilising the system to any chosen one
among them. The theoretical and experimental works so far done on
this topic shows that various methods of control are capable  of leading 
to improved performance and ordered response~\cite{ch.Boc1}.
Moreover the naturally occurring chaotic orbits on a dynamical attractor has
been utilised to carry information~\cite{ch.Kur2a}. Thus one
can encode and transmit a prescribed message using chaotic states. 
This process may offer practical advantages over the usual periodic 
carriers, such as the possibility of real time reconstructions of 
signal dropouts in the communication. In such cases, it is desirable to have
a number of systems behaving in a synchronised manner and this is  
achieved by coupling or feedback between systems.

The phenomenon of synchronisation in chaotic systems has received much
attention since 1990~\cite{ch.Pec1a}. It is found to 
occur in the dynamics of many coupled physical
\cite{ch.Tru2,ch.Tan3,ch.Ro4,ch.Tic5,ch.Are6,ch.Jun7}
and biological systems~\cite{ch.Tas8,ch.Ani9,ch.Por10} 
such as neurons in a network
\cite{ch.Ruk11,ch.Els11a,ch.Tas12,ch.Ruk13}. The synchronisation 
schemes form the basis of biological clocks that regulate
daily and seasonal rhythms of living systems from
bacteria to humans. Even when the individual systems are in the realm of chaotic behaviour, the coupled one most often
can be synchronised in periodic or regular states. Thus synchronisation
techniques automatically brings in control of chaos also. However there are
applications like information processing where synchronised chaotic
states are desirable.

There are different types of 
synchronised states~\cite{ch.Pik15,ch.Pec16} occuring in coupled 
chaotic systems. The first type called complete synchronisation has
received much attention with the development of a new type of 
communication technique that exploits the possibility of masking
a message by mixing it with a chaotic signal~\cite{ch.Pik14}.
With both the amplitudes and phases of the coupled systems 
varying in the same way the synchronised chaotic state will be 
restricted to a smooth invariant manifold of lower dimension than
the full phase space.Complete synchronisation implies that
the details of the two systems are identical or their
difference converges to zero. \cite{ch.Boc1,ch.Mur17,ch.Mur18,pc97c}. 
However there are situations where one observes lag synchronisation
in which the system states coincide only when one of them is shifted in time.
So also there are cases of generalised partial 
synchronisation where no amplitude synchronisation is observed 
but the average amplitude  is bounded and varies periodically~\cite{ch.Olg19}.

The majority of the research works reported in the literature have been 
focused on 
synchronisation of continuous systems and a few on discrete systems concentrate
on one dimensional map \cite{ch.Pik14,ch.Mur17,ch.Mur18}. 
In this paper we consider two dimensional discrete systems and introduce 
different coupling schemes  and the resulting possible types of 
synchronisation. Such systems model many interesting 
situations like population dynamics of mutually dependent species,
neural networks, image processing techniques etc. Hence the possibility of 
their evolution in a synchronised 
manner under different types of connectivity forms a useful study. 
Moreover, two dimensional discrete systems occur as the Poincar\'e maps
of three dimensional continuous systems, which are the lowest dimensional
systems capable of exhibiting chaos. Hence the synchronised behaviour
of the latter  can be analysed in a general and systematic way using the 
former.

The paper is organized as follows. 
In \S\;\ref{ch.sec2}, we describe the different coupling schemes and
discuss the stability of synchronisation for each one.
In \S\;\ref{ch.sec3} and \S\;\ref{ch.sec4} these coupling schemes are 
applied to two specific examples of discrete systems \textit{viz.} 
Gumowski-Mira map (GM) and Modified Gumowski-Mira map (MGM) maps and their 
synchronisation is studied 
using similarity function, stability analysis, synchronisation 
and stabilisation times and available basin in the parameter as well as 
initial value plane. It is found that complete synchronisation with 
control, lag synchronisation and synchronised chaotic states can be realised by choosing different coupling techniques.
Concluding remarks are given in \S\;\ref{ch.sec5}. 

\section{Coupling schemes for 2-d maps}\label{ch.sec2}
We start with a two dimensional discrete system which is written as
\begin{equation}
\overline{X1}=\overline{F}(\overline{X1},\mu)\mbox{ and }
\overline{X2}=\overline{F}(\overline{X2},\mu).
\end{equation}
where $\overline{X1}=(X1,Y1)$ define two dimensional space of the first map,
$\overline{X2}=(X2,Y2)$ the two dimensional space of the second map and 
$\mu$ is the control parameter.
When both the systems are coupled the dynamics of the system develops as
\begin{eqnarray}
\overline{X1}&=&\overline{F}(\overline{X1},\mu)+I_1(\epsilon,\overline{X1}, 
\overline{X2})\nonumber\\
\overline{X2}&=&\overline{F}(\overline{X2},\mu)+I_2(\epsilon,\overline{X1}, 
\overline{X2})
\end{eqnarray}
Here $I_1$ or $I_2$ is the coupling scheme applied to 
both the maps which in general is a function of the variables of both maps
and the coupling parameter $\epsilon$. Different types of $I$ can 
be chosen to achieve the required  synchronised state.

The possible forms for the coupling scheme $I$ that we consider in this work
are\\
(i) linear coupling (CS1) applied to one of the equations, \textit{ie.}
$$
I_i =\epsilon X_j
$$
(ii) linear difference coupling (CS2), $I_i=\epsilon(X_j-X_i)$ \\
(iii) additive parametric coupling (CS3), $I_i=\mu(1-\epsilon X_j)$.\\
For all the above cases $i,j=1,2$ and $i\neq j$.

The first two are well studied schemes of synchronisation while the third 
one is introduced for the particular type of two dimensional system 
discussed here.  
It is interesting to note that this coupling depends on the parameter $\mu$
of the individual system in addition to the usual coupling parameter 
$\epsilon$. We find that this introduces more flexibility in the scheme by 
providing two parameters, one of which controls the dynamics also. This can
result in a wider range in the  $(\mu, \epsilon)$ plane for optimum 
synchronisation. The stability analysis is carried out and the similarity 
function plotted against time to characterise the nature and robustness of 
the synchronised state.
In each case the efficiency of the scheme is quantified by 
computing the average time $\tau_1$ taken for achieving synchronisation over
a number of initial conditions. The robustness is checked by perturbing the 
synchronised state with a random noise and computing the stabilisation time 
$\tau_2$ required to regain synchronisation after perturbation.

Regions in the parameter plane $(\mu, \epsilon)$ that leads to different
synchronised states are isolated and the basin of synchronisation in the phase
plane marked out. These characterisations help to estimate the relative
merits of different schemes.

 The stability of the symmetric state
$X_i=X_j$ and $Y_i=Y_j$ with respect to the transverse perturbations
$(\delta\overline{X},\delta\overline{Y})$ is analysed by linearisation. We get
\begin{eqnarray}
\delta\overline{X1} &=& \frac{\partial F}{\partial X1} \delta X1+
\frac{\partial F}{\partial Y1} \delta Y1+
\epsilon \overline{C} (\delta\overline{X2}, \delta\overline{X1})\nonumber\\
\delta\overline{X2} &=& \frac{\partial F}{\partial X2} \delta X2+
\frac{\partial F}{\partial Y2} \delta Y2+
\epsilon \overline{C} (\delta\overline{X1}, \delta\overline{X2})
\end{eqnarray}
where $\overline{C}$ is in general a function of $\delta\overline{X1}$, 
$\delta\overline{X2}$. Defining $\overline {V}= (\delta \overline{X}_1 - 
\delta \overline{X}_2)$ and $F_{ij}=\frac{\partial F_i}{\partial X_j}$.

(i) For linear coupling this is obtained as
\begin{equation}
V^i_{n+1}=\sum\limits_j F_{ij}V_n^j-\epsilon C_i \mbox{ with }
C_1=1, C_2=0, \;i=1,2.
\end{equation}

(ii) For linear difference coupling, 
\begin{equation}
V^i_{n+1}=\sum\limits_j F_{ij}
V^j_{n}-2\epsilon C_i 
\end{equation}
with $C_1=1$ and $C_2=0$.

(iii) For additive parametric coupling,
\begin{equation}
V^i_{n+1}=\sum\limits_j F_{ij}V^j_{n}+\mu\epsilon C_i \mbox{ with }
C_1=1, C_2=1.
\end{equation}

Eigen values $\sigma^i$ of the Jacobian matrix $F_{ij}$  are evaluated
and the transverse Lyapunov exponents $(\lambda_\perp)$ calculated.\\
For CS1, 
\begin{equation}
\lambda_\perp^i = \ln (\sigma^i - \epsilon).
\end{equation}
For CS2, 
\begin{equation}
\lambda_\perp^i = \ln (\sigma^i - 2\epsilon).
\end{equation}
For CS3, 
\begin{equation}
\lambda_\perp^i = \ln (\sigma^i +\mu \epsilon).
\end{equation}
The largest transverse Lyapunov exponent being less than zero
indicates the stability of the synchronised state \cite{ch.Pik14}.

\section{Synchronisation in Gumowski-Mira maps}\label{ch.sec3}

As a first example of a two dimensional discrete map 
we consider Gumowski-Mira Map (GM)\cite{ch.Gum21}. 
This is a two dimensional difference
equation in $(X,Y)$ involving a non linear function $f(X)$ that
is advanced in time by one iteration in the $Y$ equation.
The iterates of this map give rise to a variety of interesting
patterns in the $(X,Y)$ plane and hence useful as a two dimensional
iterative scheme for producing fractal objects.
The GM transformation is defined in the $(X,Y)$ plane as
\begin{eqnarray}
X_{n+1}             &=& Y_{n}+a(1-bY_n^2)Y_n+f(X_n)\nonumber\\
Y_{n+1}             &=& -X_n+f(X_{n+1})\label{ch.eq1}\\
{\rm with }\quad f(X_n)  &=& \mu X_n+\frac{2(1-\mu)X^2_n}{1+X_n^2}.\nonumber
\end{eqnarray}

\begin{figure}
\centering
\mbox{%
\epsfig{file=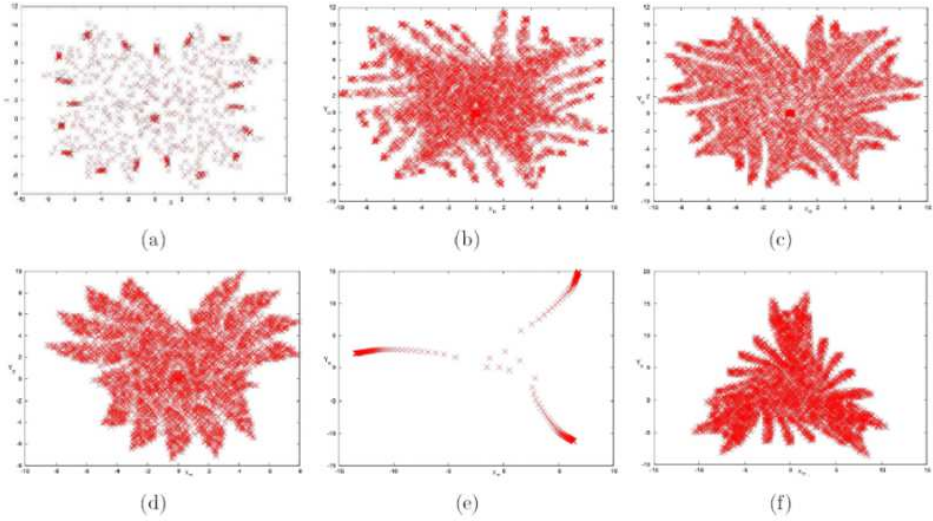,width=.8\textwidth}}
\caption{GM patterns generated in the phase space ie. $(X,Y)$ plane 
by the discrete system in equation (\ref{ch.eq1}). Here $a=0.008$, $b=0.05$, 
$X=0.1$, $Y=0.0$. $-0.6\leq \mu\leq -0.1$, $\Delta\mu=0.1$. It is clear that 
the asymptotic dynamic states depends very sensitively on the control 
parameter $\mu$.}\label{ch.fig1}
\end{figure}

This map is found to have a very sensitive dependence on the control parameter
$\mu$ \cite{ch.Ots22,ch.Amb26}. The phase space plots known as GM patterns in 
Fig. (\ref{ch.fig1}a) justify this.
In these calculations  $a$ and  $b$ are fixed to be
$0.008$ and $0.05$ respectively and $\mu$ is increased in steps of 0.1.  
In an earlier analysis we have found that 
many recurring periodic and self similar sub structures, each
with its own intermittency, periodicity, quasi periodic bands,
band merging etc. in the bifurcation scenario of the 
system \cite{ch.Amb26} explain the dynamics of these patterns.

Here we apply different coupling schemes mentioned in the previous section 
to the 
above system and try to target it to lower stable periodicities in their
synchronised states.

\begin{figure}
\centering
\mbox{
\epsfig{file=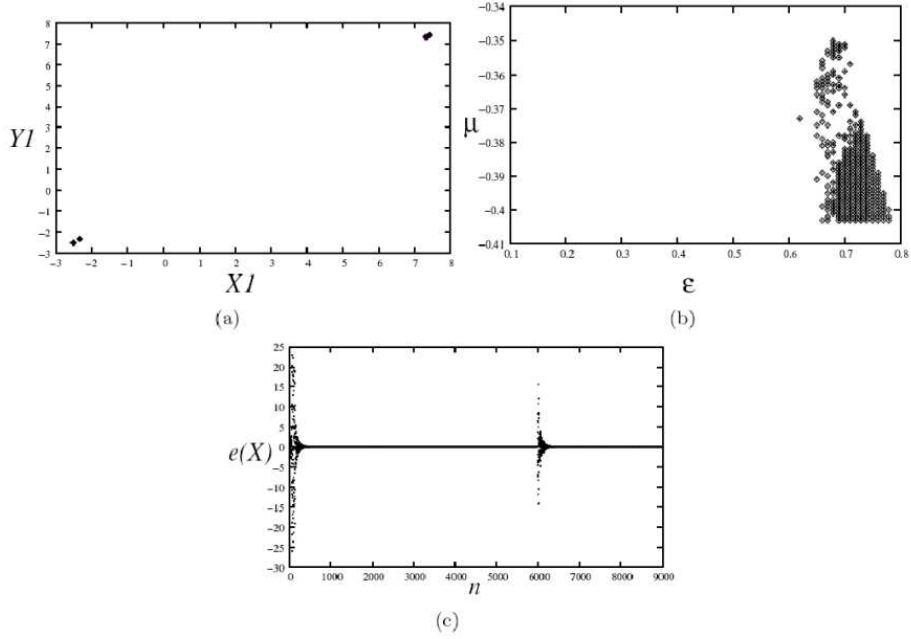,width=.8\textwidth}}
\caption{Linear coupling applied to two GM maps
a) The synchronised four cycle for $\mu=-0.39$. b) The points in the 
$(\mu,\epsilon)$ plane that lead to synchronised four cycle.
c) The error function $e(X)$ ploted against iteration time $n$. When a random
perturbation is applied after 6000 iterations synchronisation is regained in
410 steps.}\label{ch.fig2}
\end{figure}

In the case of CS1  the pair of synchronising systems evolve as
\begin{eqnarray}
X1_{n+1} &=& Y1_n +a(1-bY1_n^2)Y1_n+f(X1_n)+\epsilon X2_n\nonumber\\
Y1_{n+1} &=& -X1_n+f(X1_{n+1})\label{ch.eq2}\\
X2_{n+1} &=& Y2_n +a(1-bY2_n^2)Y2_n+f(X2_n)+\epsilon X1_n\nonumber\\
Y2_{n+1} &=& -X2_n+f(X2_{n+1})\nonumber
\end{eqnarray}
where $\epsilon$ is the coupling parameter. The advantage of the method is that
the feedback in the $X$ equation alone suffices to achieve total
synchronisation of both the variables. We start with  $\mu=-0.39$ 
(the individual systems are in the chaotic region), for $\epsilon=0.7$ and 
initial conditions $(0.1,0.0), (0.2,0.12)$, the $X$ and $Y$ equations of both settle to an 
identical four  cycle which is clear from Fig.(\ref{ch.fig2}a).
Fig (\ref{ch.fig2}b)
gives the $(\mu,\epsilon)$ plot for synchronising values.  
The error function defined  as $e(X)=X1-X2$ and 
$e(Y)=Y1-Y2$ tends to zero when synchronisation is achieved \cite{ch.Vin27}.
We test this numerically by iterating equation (\ref{ch.eq2}) 
for values mentioned above and plotting $e(X)$ against iteration number 
$n$ (Fig (\ref{ch.fig2}c)).
The synchronisation time $\tau_1$ and stabilisation time $\tau_2$ 
which is the time taken to stabilise after applying a random noise 
for 10 iterations are
shown in the figure. The values of $\tau_1$ and $\tau_2$ are found for
10 different initial conditions and their average calculated. We get
$\tau_1=778$ and $\tau_2=410$. 
 The largest transverse Lyapnunov exponent for the four cycle
works out to be $-0.1564$. This indicates that this synchronised 
state is stable. 

However this coupling  is active even after achieving synchronisation.
Hence to make the scheme cost effective,we try a 
linear difference coupling (CS2) of the two systems,
\begin{eqnarray}
X1_{n+1} &=& Y1_n +a(1-bY1_n^2)Y1_n+f(X1_n)+\epsilon(X2_n-X1_{n})\nonumber\\
Y1_{n+1} &=& -X1_n+f(X1_{n+1})\label{ch.eq3}\\
X2_{n+1} &=& Y2_n +a(1-bY2_n^2)Y2_n+f(X2_n)+\epsilon (X1_n-X2_n)\nonumber\\
Y2_{n+1} &=& -X2_n+f(X2_{n+1})\nonumber
\end{eqnarray}

With  $\mu=-0.23$ both the systems are individually chaotic.
When $\epsilon=0.48$ and initial conditions $(0.1,0.0), (0.2, 0.12)$  they 
stabilise to an eight cycle, with a time lag $\tau=4$ between them.
Synchronisation achieved here is lag synchronisation which can be 
confirmed using similarity function, defined as \cite{ch.Ros24}
\begin{equation}\label{ch.add-eq4}
S_l^2(\tau)=
\frac{\big\langle [X2_{n+\tau}-X1_n]^2\big\rangle}
{\big[\langle X1_n^2\rangle\langle X2_n^2\rangle \big]^{1/2}}
\end{equation}
For a non zero value of $\tau$, $S_l(\tau)\rightarrow 0$
corresponds to lag synchronisation. 
\begin{figure}
\centering
\mbox{
\epsfig{file=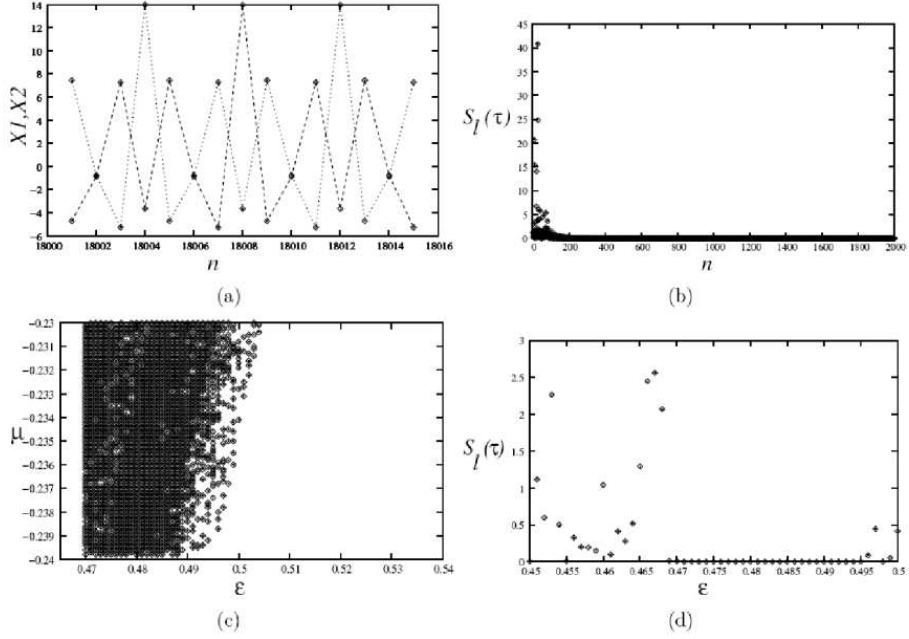,width=.8\textwidth}}
\caption{Lag synchronisation observed in GM maps under Linear difference coupling
a) $(X1,X2)$ plot with iteration number $n$ showing synchronisation with a lag
of $\tau=4$. b) Variation of similarity function with iteration number $n$ for
$\tau=4$. c) Regions in the $(\mu,\epsilon)$ plane that can lead to lag
synchronisation. 
d) Variation of similarity function as a function of coupling 
parameter $\epsilon$ plotted after 10000 steps for $\mu=-0.23$. Lag 
synchronisation is realised only for a range of $0.471<\epsilon<0.495$.}\label{ch.fig3}
\end{figure}
It is clear from 
Fig (\ref{ch.fig3}a),  the time series plot of $(X1_n,X2_n)$
with $\tau=4$. Fig (\ref{ch.fig3}b) gives the $S_l(\tau)$ vs $n$ plot which 
establishes lag synchronisation phenomena in the system. Fig (\ref{ch.fig3}c) 
gives the  $(\mu,\epsilon)$ plot for synchronising values. 
Fig (\ref{ch.fig3}d) gives the variation of similarity function
$S_l(\tau)$ with $\epsilon$. 
It is seen that lag synchronisation occurs only in a narrow region of 
$\epsilon$ and the states are de-synchronised in the remaining regions.
The largest transverse Lyapunov exponent, 
$\lambda_\perp=-0.431144$, for the state with lag synchronisation.

The  synchronisation in this scheme is only lag type.
To achieve total synchronisation in a periodic state
we try an additive parametric coupling (CS3).

The dynamics of the coupled GM maps on applying CS3
 develops through the equations
\begin{eqnarray}
X1_{n+1} &=& Y1_n +a(1-bY1_n^2)Y1_n+f(X1_n)+\mu(1-\epsilon X2_n)\nonumber\\
Y1_{n+1} &=& -X1_n+f(X1_{n+1})+\mu (1-\epsilon Y2_n)\label{ch.eq4}\\
X2_{n+1} &=& Y2_n +a(1-bY2_n^2)Y2_n+f(X2_n)+\mu(1-\epsilon X1_n)\nonumber\\
Y2_{n+1} &=& -X2_n+f(X2_{n+1})+\mu (1-\epsilon Y1_n).\nonumber
\end{eqnarray}
For $\mu=0.001$, both the maps individually exhibit
irregular and aperiodic behaviour. 
\begin{figure}
\centering
\mbox{
\epsfig{file=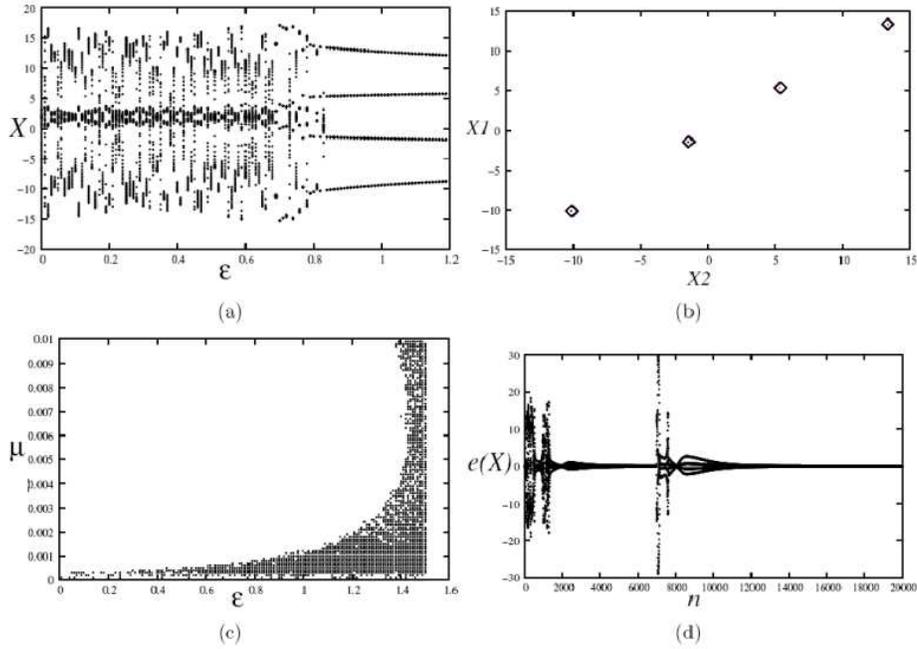,width=.8\textwidth}}
\caption{Additive parametric coupling  applied in two GM maps for
$\mu=0.001$ a) $\epsilon$ vs $X1$ plot; the two system settles to four cycle 
for $\epsilon=0.8$ indicating control of chaos before the onset of 
synchronisation; 
b) The synchronised periodic four cycle as $\epsilon$ is increased to $0.85$;
c) The values in the $(\mu,\epsilon)$ plane that leads to synchronisation in 
the periodic four cycle;
d) The evolution of the error function $e(X)$ showing complete synchronisation
and return to synchronisation after applying the random perturbation at the 
7000$^{\rm th}$ step.}\label{ch.fig4}
\end{figure}
After coupling with
$\epsilon=0.85$ and initial values $(0.1, 0.0)$, $(0.2, 0.12)$ they are found 
to settle to the same four cycle. Fig (\ref{ch.fig4}a)
gives the $(\epsilon,x)$ plot for the first map which shows that for
$\epsilon=0.8$ the map settles to a four cycle \textit{ie.} control
of chaos is taking place before synchronisation and synchronisation is 
reached only when $\epsilon$ is increased to 0.85. 
Fig (\ref{ch.fig4}b) gives
the $(X1,X2)$ plot showing synchronisation and in fig (\ref{ch.fig4}c)  the
values of $\mu$ and $\epsilon$ for which  synchronisation is achieved are
shown.

The variation of the error function is calculated 
numerically by iterating equation (\ref{ch.eq4}) and is plotted against
iteration number $n$ in Fig (\ref{ch.fig4}d). 
The average time taken for synchronisation is found
to be $\tau_1=6265$ and the average stabilisation time $\tau_2=3844$. 
The largest transverse Lyapnunov exponent is calculated as explained  in the 
previous case and is obtained as $\lambda_\perp=-2.75098$. Thus the 
synchronised state is a periodic four cycle which is stable.

\begin{figure}
\centering
\mbox{%
\epsfig{file=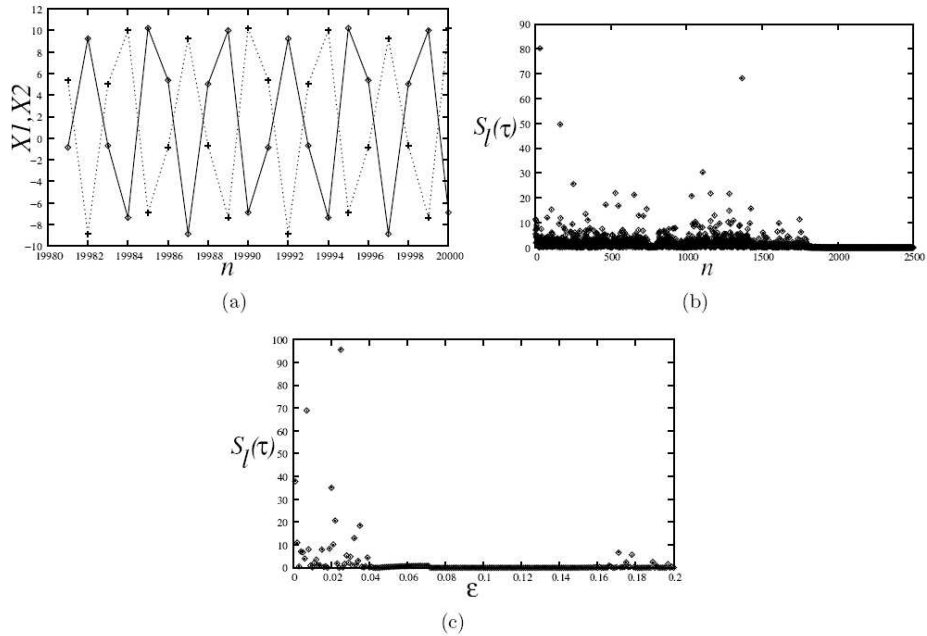,width=.8\textwidth}}
\caption{The lag synchronisation observed under additive parametric coupling 
for $\epsilon=0.1$, $\mu=-0.3$ with an asymptotic 10 cycle 
(a) $X1$ and $X2$ ploted with iteration number $n$ showing a lag of $\tau=5$;
(b) the variation of the similarity function to zero after $n=1800$
(c) the variation of similarity function as a function of coupling parameter
$\epsilon$. Lag synchronisation is realised in the range 
$0.075<\epsilon<0.15$.}\label{ch.fig5}
\end{figure}

With $\mu=-0.3$, initial conditions $(0.1,0), (0.2, 0.12)$ and
$\epsilon=0.1$ both the systems are found to synchronise to a
stable ten cycle. We note that this is also a case of lag synchronisation \textit{ie.}, $X1_n=X2_{n+\tau}$,with $\tau=5$.
The time series plot of $X1_n$ and $X2_n$ in Fig (\ref{ch.fig5}a)
and $S_l(\tau)$ vs $n$ plot in Fig (\ref{ch.fig5}b) establishes
this phenomenon in the system. We observe lag synchronisation
for the range $0.075 <\epsilon < 0.15$ (Fig (\ref{ch.fig5}c)).  

The range over which lag synchronisation exists in this case is larger than
the previous type of coupling. However beyond $\epsilon=0.15$, the two systems
are totally desynchronised.

\section{Synchronisation in Modified Gumowski-Mira maps}\label{ch.sec4}
As a second example we consider a variant of the GM map which 
is hereafter called modified GM map (MGM map), defined by the following set 
of equations.
\begin{eqnarray}
X_{n+1} &=& Y_n+a(1-bY^2_n)Y_n+f(X_n)\nonumber\\
Y_{n+1} &=& -X_n+f(X_n)\\
\mbox{with }\;f(X_n) &=& \mu X_n +\frac{2(1-\mu)X^2_n}{1+X_n^2}.\nonumber
\end{eqnarray}
In this map the iterations in both $X$ and $Y$ are in step.
With values of $a$ and $b$ are fixed as $a=0.008$, $b=0.05$,
the system has a prominent  period doubling sequence in the bifurcation 
scenario \cite{ch.Amb23}. In the GM map considered in \S\;\ref{ch.sec3} 
above synchronised states
are periodic states which means synchronisation follows control of chaos. We
find that synchronised chaotic states can be realised in addition to 
synchronised fixed points and limit cycles in the modified GM map
introduced here.

We apply CS1 to two such systems.
In this case for $\mu=-0.7$ and initial conditions $(0.1,0),(0.2,0.1)$ both the
maps are individually chaotic. With coupling for $\epsilon=0.4$ they
asymptotically approach a fixed point $(1.66,-0.3316)$, while for 
$\epsilon=0.5$ they synchronise to a limit cycle.

\begin{figure}
\centering
\mbox{%
\epsfig{file=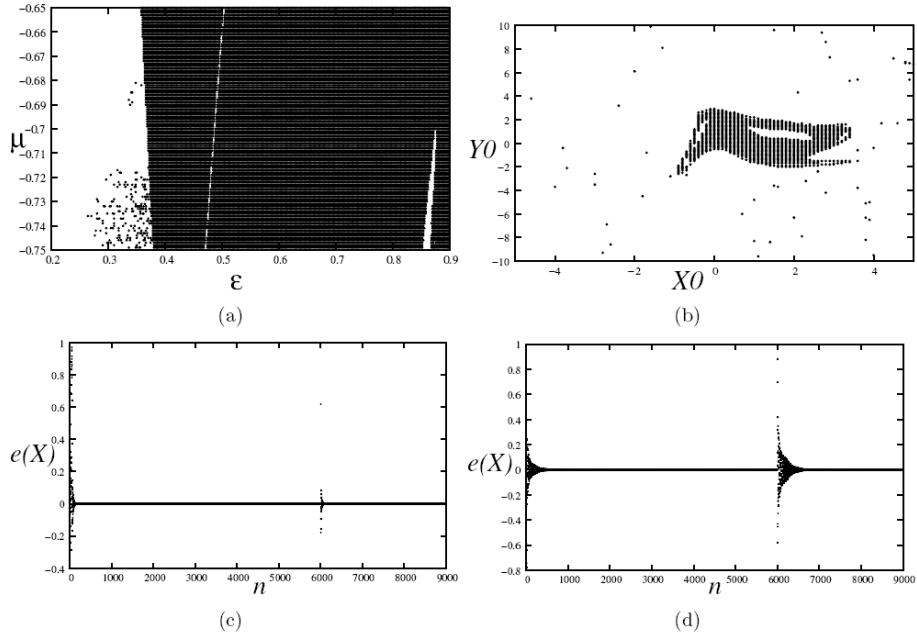,width=.8\textwidth}}
\caption{Linear coupling applied to two MGM maps 
a) Regions in $(\mu,\epsilon)$ plane leading to synchronised states. 
b) basin of the synchronised states in the initial value plane.
c) error function $e(X)$ vs iteration number $n$ plot for $\epsilon=0.4$ where
the synchronised state is a fixed point. A random noise is applied at the 
6000$^{\rm th}$ iterate. Synchronisation is regained after 98 iterations. 
d) error function $e(X)$ vs $n$ plot for $\epsilon=0.5$ where the synchronised
state is a limit cycle. A random noise is applied at the 6000$^{\rm th}$ 
iterate. Synchronisation is regained after 520 iterations.}\label{ch.fig6}
\end{figure}

Fig (\ref{ch.fig6}a) gives the $(\mu,\epsilon)$ plot where synchronisation 
sets 
in. 
The region to the left of the white line corresponds to synchronisation
to a fixed point while that to the right corresponds to limit cycle.
Fig (\ref{ch.fig6}b) gives the basin of synchronisation of two 
MGM maps for
$\epsilon=0.5$ (when the final synchronised state is a limit cycle).
Fig (\ref{ch.fig6}c) and Fig (\ref{ch.fig6}d) gives the $e(X)-n$ plot of 
coupled MGM maps from where synchronisation time and stabilisation time can be
read off.  For
$\mu=-0.7$ and $\epsilon=0.4$ the largest  transverse Lyapunov exponent
$(\lambda_\perp)$ is $(-1.883383)$ where one cycle periodicity is seen in the
synchronised state. For $\epsilon=0.5$ when the synchronised state is a limit
cycle the value $\lambda_\perp$ is $(-1.44478)$.

\begin{figure}
\centering
\mbox{%
\epsfig{file=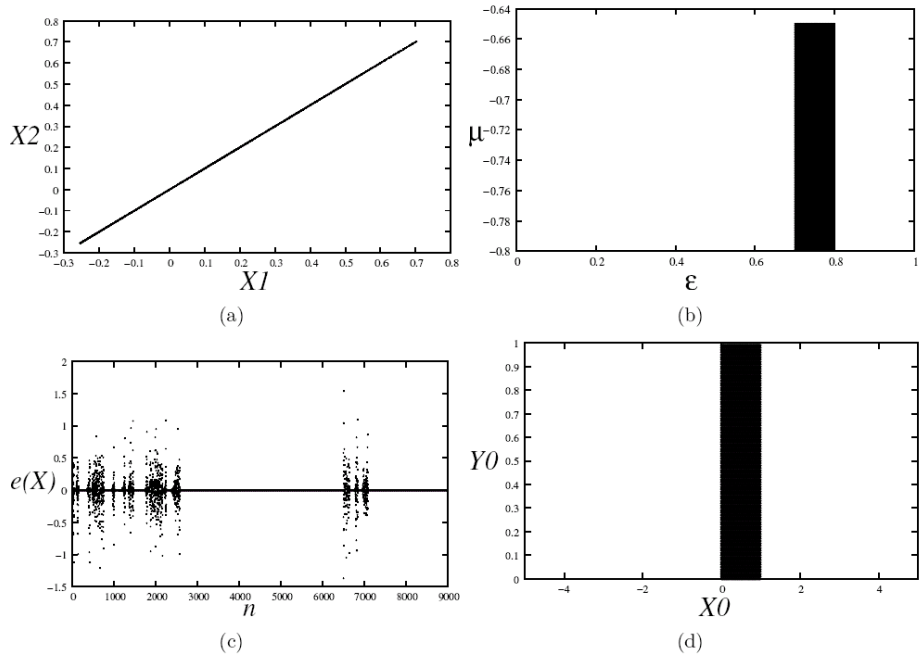,width=.8\textwidth}}
\caption{Linear difference coupling applied to two MGM maps.
a) $(X1,X2)$ plot showing synchronised chaotic states
b) regions in the $(\mu,\epsilon)$ plane where synchronisation occurs.
c) Variation of the error function $e(X)$ with $n$ for $\mu=-0.7$. The time 
taken to reach synchronisation and the time taken to regain the same after a
noisy perturbation are more in this case
d) basin of synchronised chaotic state in the initial value plane.}\label{ch.fig7}
\end{figure}

We try CS2 for two MGM maps.
For $\mu=-0.7$ both the systems are individually chaotic and for 
$\epsilon=0.85$ we observe chaotic states synchronised in the coupled system. 
This
is shown in $(X1,X2)$ plot in Fig (\ref{ch.fig7}a). Fig (\ref{ch.fig7}b) gives 
the regions in the $(\mu,\epsilon)$ plane leading to this type of 
synchronisation.
Error function $e(X)$ and $e(Y)$ evolve into a stable fixed point
$(0,0)$, which is shown in Fig (\ref{ch.fig7}c). The average synchronisation 
time and stabilisation time for 10 different initial conditions are found 
to be $\tau_1=2605$ and $\tau_2=1057$. Fig (\ref{ch.fig7}d) gives the basin for synchronisation in
the range [-5,5]. $\lambda_\perp=-0.40786$ shows  that the chaotic
synchronised state is stable.

We try CS3 scheme to two MGM maps.

\begin{figure}
\centering
\mbox{%
\epsfig{file=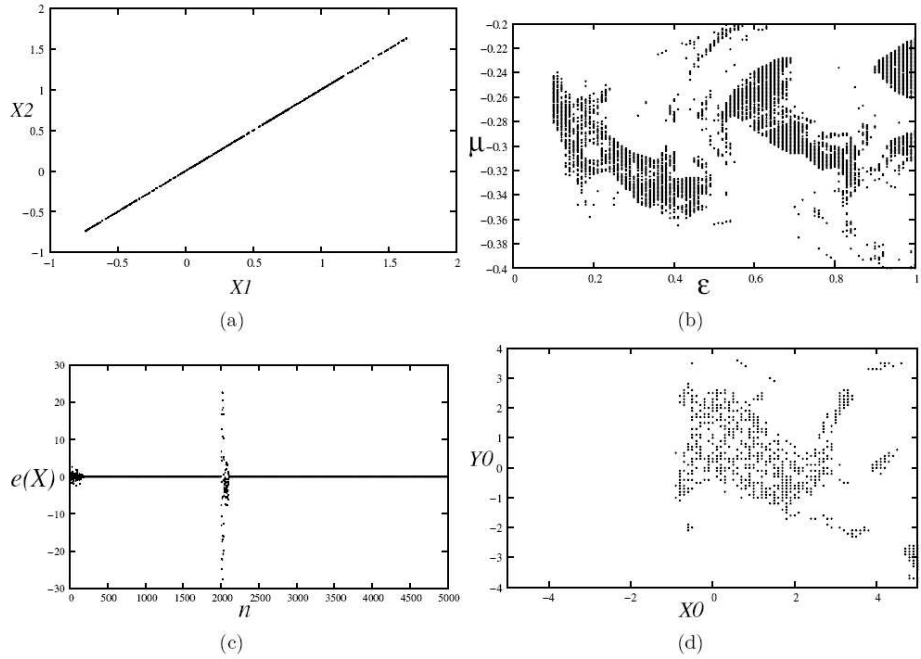,width=.8\textwidth}}
\caption{Additive parametric coupling in two MGM maps
a) The $(X1,X2)$ plot of the synchronised chaotic state 
b) points in the $(\mu,\epsilon)$ plane resulting in synchronisation.
c) Variation of error function $e(X)$ with time for $\mu=-0.3$. 
d) The basin of synchronised chaotic state in the initial value plane.
This scheme is more efficient and the state is robust to 
perturbation.}\label{ch.fig8}
\end{figure}

Here for $\mu=-0.3$ initial conditions $(0.1, 0.0) (0.2, 0.1)$
(where individual systems show 4 cycle periodicity) for
$\epsilon=0.5$, we observe synchronised chaotic states in the
coupled systems. This is shown in $(X1,X2)$ plot in Fig (\ref{ch.fig8}a).
Fig (\ref{ch.fig8}b) gives the regions of the parameter plane
$(\mu,\epsilon)$ leading to this type of synchronisation.

$e(X)$ evolve into a stable fixed point $(0,0)$ which is shown in 
Fig (\ref{ch.fig8}c). A similar plot is obtained for $e(Y)$ also. 
Time taken for
synchronisation $\tau_1=282$. After 2000 iterations a noise is applied and 
stabilisation time is obtained as $\tau_2=186$. Fig (\ref{ch.fig8}d) gives the
basin for synchronisation in the range [-5,5].

For this coupling,
the largest transverse Lyapunov exponent computed works out to be 
$-1.227154$ indicating stability for synchronised chaotic state.

\section{Conclusion}\label{ch.sec5}

In this work, a study on different coupling schemes for synchronisation 
in two dimensional systems is carried out. We consider two types of two dimensional discrete
maps called Gumowski-Mira(GM) maps and modified Gumowski-Mira(MGM) maps. Three
different coupling schemes namely linear coupling(CS1), linear difference coupling(CS2)
and additive parametric coupling(CS3) are applied to them. 
In GM maps using  CS1 and CS3 control of chaos is achieved along with 
complete synchronisation. However only lag synchronisation is  observed with 
CS2. In MGM maps while CS1 stabilises the maps to fixed points or limit 
cycles, CS2 and CS3 lead to synchronised chaotic states. It is found that the 
synchronisation time and stabilisation time is shortest for CS1 in GM maps 
compared to CS3. In the case of MGM maps also CS1 offers the shortest 
synchronisation time and stabilisation time. In this case both CS2 and CS3  
yields chaotic synchronised states among which $\tau_1$ and $\tau_2$ are 
shorter for CS3.
It is found that the basin of synchronisation in the initial value plane 
is unique for all the three types of coupling in GM maps, 
while the regions leading to synchronisation in MGM maps are small but 
finite. The parameter values $(\mu, \epsilon)$ leading to synchronisation 
in GM maps has largest range for CS3 while those leading to lag 
synchronisation in CS2 has limited range. In MGM maps this range is maximum 
for CS1 and minimum for CS2.
It is found that although CS2 is cost effective, CS1 gives maximum 
range in $(\mu,\epsilon)$ plane, wider basin in $(X0,Y0)$ plane and 
shorter times for synchronisation and stabilisation after perturbation. 
CS3 has wider range in parameter with control of chaos in GM map and is 
useful to get synchronised chaotic states in MGM maps.

Although control of chaos is simultaneous with synchronisation in most
cases, synchronised chaotic states are possible in the second example 
considered here. This may find applications in information processing using
connected systems. 

\subsection*{Acknowledgement}
K. A. thanks University Grants Commission, New Delhi for deputation
under Faculty Improvement Programme.

\end{document}